\documentclass[aps,prl,twocolumn,showpacs,superscriptaddress]{revtex4-1}

\usepackage{amsfonts}
\usepackage{amsmath}
\usepackage{amssymb}
\usepackage{graphicx}
\usepackage{bm}

\begin{document}



\title{ Anisotropic resistivity in underdoped single crystals (Ba$_{1-x}$K$_x$)Fe$_2$As$_2$, $0 \leq x<0.35$}


\author{M.~A.~Tanatar}
\email[Corresponding author: ]{tanatar@ameslab.gov}
\affiliation{Ames Laboratory, Ames, Iowa 50011, USA}

\author{E.~C.~Blomberg}

\affiliation{Ames Laboratory, Ames, Iowa 50011, USA}
\affiliation{Department of Physics and Astronomy, Iowa State University, Ames, Iowa 50011, USA }

\author{Hyunsoo Kim}

\affiliation{Ames Laboratory, Ames, Iowa 50011, USA}
\affiliation{Department of Physics and Astronomy, Iowa State University, Ames, Iowa 50011, USA }

\author{Kyuil Cho}

\affiliation{Ames Laboratory, Ames, Iowa 50011, USA}
\affiliation{Department of Physics and Astronomy, Iowa State University, Ames, Iowa 50011, USA }

\author{W. E. Straszheim}
\affiliation{Ames Laboratory, Ames, Iowa 50011, USA}

\author{ Bing Shen }
\affiliation{ Institute of Physics, Chinese Academy of Sciences, Beijing 100190, P. R. China }

\author{Hai-Hu Wen}
\affiliation{ Institute of Physics, Chinese Academy of Sciences, Beijing 100190, P. R. China }
\affiliation{ National Laboratory of Solid State Microstructures and Department of Physics, Nanjing University, Nanjing 210093, P. R. China }
\affiliation{Canadian Institute for Advanced Research, Toronto, Ontario, Canada}

\author{R.~Prozorov}
\affiliation{Ames Laboratory, Ames, Iowa 50011, USA}
\affiliation{Department of Physics and Astronomy, Iowa State University, Ames, Iowa 50011, USA }

\date{\today}


\begin{abstract}

Temperature-dependent in-plane, $\rho _a(T)$, and inter-plane, $\rho _c(T)$, resistivities were measured for the iron-arsenide superconductor (Ba$_{1-x}$K$_x$)Fe $_2$As$_2$ over a broad doping range from parent compound to optimal doping $T_c\approx 38~K$, $0\leq x \leq 0.35$. The coupled magnetic/structural transition at $T_{SM}$ is clearly observed for samples with $T_c <$26~K ($x <0.25$), however its effect on resistivity is much weaker than in the electron-doped Ba(Fe$_{1-x}$Co$_x$)Fe $_2$As$_2$, and the transition leads only to a decrease of resistivity. In addition to the feature at $T_{SM}$, the inter-plane resistivity shows a maximum at $T^*\sim$200~K, which moves slightly to higher temperature with doping, revealing a trend opposite to the electron-doped materials. A smeared feature at about the same temperature is seen in $\rho _a(T)$. For $T<T^*$, the temperature dependence of resistivity shows systematic evolution and is close to linear at optimal doping. This feature, being most pronounced for $\rho_c(T)$, suggests the existence of a quantum critical point close to optimal doping.

\end{abstract}

\pacs{74.70.Dd,72.15.-v,74.25.Jb}




\maketitle



\section{Introduction}

Superconductivity in hole-doped (Ba$_{1-x}$K$_x$)Fe$_2$As$_2$ \cite{Rotter} (BaK122, in the following) was found soon after the discovery of superconductivity with high critical temperatures in oxypnictide FeAs-based materials \cite{Hosono}.
Several studies reported anisotropic properties of single crystals, representative of various parts of the phase diagram \cite{NiNiK,Welp,Wencrystals,Chen1,Yuan,Shiyan,Terashimaresistivity,HashimotoKpure}, however, no systematic evolution of the resistivity and its anisotropy was undertaken so far, due to a difficulty in preparation of single crystals with well controlled potassium content.

High quality single crystals of K-doped materials can be grown from FeAs flux \cite{Wencrystals}, however, the high melting temperature of the flux, leading to high potassium vapor pressure, limits this technique to growth of only the underdoped compositions. The compositions on the overdoped side can be grown from KAs flux \cite{KAsfluxgrowth}. Initial success in growing single crystals through the use of Sn flux, and finding their low anisotropy through measurements of the upper critical field \cite{NiNiK}, was stopped by finding of the macroscopic phase separation in Sn-flux \cite{Kphaseseparation,Kphaseseparation2,Kphaseseparation3} or polycrystalline \cite{phaseseparationpoly} samples  and gross effect of small Sn incorporation on the phase diagram \cite{Sneffect}.

Measurements of anisotropy are of great importance for understanding the normal state of iron pnictide superconductors. For example, careful characterization of the anisotropic resistivity in electron-doped Ba(Fe$_{1-x}$Co$_x$)$_2$As$_2$ (BaCo122 in the following) found unusual anisotropy of transport both in the normal \cite{anisotropy,anisotropypure,detwinning,pseudogap,Fisher} and in the superconducting \cite{TanatarPRL,ReidCo} states. The in-plane transport reveals close to linear $\rho_a(T)$ at optimal doping, which evolves systematically towards $T^2$ behavior in the heavily overdoped compositions, suggestive of a quantum critical point at optimal doping \cite{NDL}. The inter-plane resistivity, $\rho_ c(T)$, on the other hand, reveals an unusual maximum \cite{anisotropy,pseudogap}, correlating well with temperature-dependent NMR Knight shift \cite{NMRPGCo} representative of a pseudogap. The characteristic temperature of the pseudogap decreases with Co-doping and vanishes at $x$=0.31 recovering normal metallic properties, in particular suppressing temperature - dependent spin susceptibility \cite{pseudogap} and Hall effect \cite{Halljapanese,HallWen}.

Broad crossover with notable slope change of temperature-dependent resistivity is also observed in in-plane transport in single crystals of BaK122 at doping close to optimal \cite{Zverev}, similar to pure stoichiometric KFe$_2$As$_2$ (K122) \cite{Shiyan,Terashimaresistivity,HashimotoKpure}. It was suggested that this unusual temperature dependence of resistivity stems from multi-band effects \cite{Zverev}, with the contribution of conductivity channels with nearly temperature-independent and strongly temperature - dependent resistivities. On the other hand, multi-component analysis of in-plane resistivity $\rho _a (T)$ in isoelectron Ru-doped BaRu122, suggests that a crossover-type temperature dependence is characteristic of hole contribution, while electronic contribution is close to $T$-linear \cite{Alloul}.

In this article we perform detailed study of the temperature-dependent in-plane and inter-plane resistivity of BaK122 over a broad doping range from parent compound to close to optimal doping level $x_{opt} \sim 0.4$ \cite{RotterPhaseD,Avci}.
We show that the unusual temperature dependence of the in-plane resistivity correlates with the pseudo-gap resistivity maximum in the inter-plane resistivity. This is dramatically different from the lack of any pseudo-gap related features in the temperature-dependent in-plane resistivity of electron-doped materials. Another difference between electron- and hole- doped materials is an increase of the resistivity crossover temperature $T^*$ in BaK122 with doping.

\section{Experimental}

Single crystals of BaK122 were grown using high temperature FeAs flux technique \cite{Wencrystals}.
Because of the volatility of K during growth, single crystals have a distribution of potassium content, with inner parts of the crystals frequently having $T_c$ differing by 1 to 3 K from the surface parts. For our study we selected samples using the sharpness of the superconducting transition as a measure of constant dopant concentration.

Samples for the study were cut from the inner parts of single crystals. After cutting, we performed precision magnetic susceptibility measurements so that we could inspect the samples for possible inclusions with lower $T_c$. In addition samples were extensively characterized by magneto-optic techniques to look for possible inhomogeneity, as described in detail in Ref.~\onlinecite{SUST}. Only samples with sharp transitions were selected. The chemical composition was measured on selected crystals using wavelength dispersive x-ray spectroscopy (WDS) in JEOL JXA-8200 electron microprobe. The composition was measured for 12 points per single crystal and averaged.

Samples for in-plane resistivity measurements had typical dimensions of (1- 2)$\times$0.5$\times$(0.02-0.1) mm$^3$. All sample dimensions were measured with an accuracy of about 10\%. Contacts for four-probe resistivity measurements were made by soldering 50 $\mu$m silver wires with ultrapure Sn solder, as described in Ref.~\onlinecite{SUST}. This technique produced contact resistance typically in the 10 $\mu \Omega$ range. Inter-plane resistivity was measured using a two-probe technique, relying on the negligibly small contact resistance. Samples typically had dimensions 0.5$\times$0.5$\times$0.1 mm$^3$ ($a\times b\times c$), their top and bottom ab-plane surfaces were covered with Sn solder forming a capacitor-like structure. Four-probe scheme was used down to the sample to measure series connected sample, $R_s$, and contact, $R_c$ resistance. Taking into account that $R_s \gg R_c$, contact resistance represents a minor correction of the order of 1 to 5\%. This can be directly seen for our samples for temperatures below the superconducting $T_c$, where $R_s =$0 and the measured resistance represents $R_c$ \cite{anisotropy,SUST,vortex}.
The details of the measurement procedure can be found in Refs.~\onlinecite{anisotropy,anisotropypure,pseudogap}.

The drawback of the measurement on samples with $c \ll a$ is that any inhomogeneity in the contact resistivity or internal sample connectivity admixes the in-plane component due to redistribution of the current. To minimize this effect, we performed measurements of $\rho_c$ on at least 5 samples of each composition. In all cases we obtained qualitatively similar temperature dependencies of the electrical resistivity, as represented by the ratio of resistivities at room and low temperatures, $\rho _c (0)/\rho _c (300)$. The resistivity value, however, showed a notable scatting and at room temperature was typically in the range 1000 to 2000 $\mu \Omega$~cm.

\section{Results}

\begin{figure}[tbh]%
\centering
\includegraphics[width=8cm]{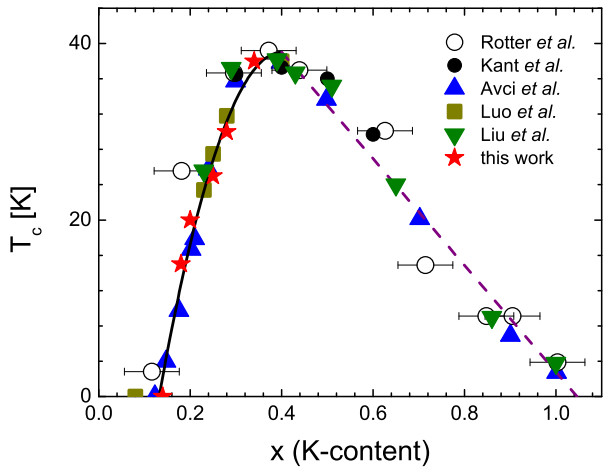}%
\caption{(Color online) Doping evolution of the superconducting transition temperature as determined from measurements on polycrystals: resistivity and magnetization \cite{RotterPhaseD} (open black circles), specific heat \cite{Rotterspecificheat} (solid black circles),  magnetization and neutron scattering \cite{Avci} (solid blue up-triangles). Dark-yellow solid squares show the data for single crystals \cite{Wencrystals}. Green down-triangles show WDS measurements on single crystals \cite{Liu}. Red stars show $T_c(x)$ as determined in this study, with $T_c$ determined from zero resistance criterion and $x $ from WDS measurements for the representative crystals. Solid black line shows a fit of the $T_c(x)$ used to determine $x$ for individual crystals in this and related \cite{BaKpenetrationdepth,BaKthermalconductivity} studies. Dashed purple line is a fit of $T_c(x)$ in the overdoped regime. }%
\label{f1_Tcx}%
\end{figure}

Fig.~\ref{f1_Tcx} shows superconducting $T_c$ as a function of potassium content. Here we compare our measurements on representative single crystals with WDS K-content determination with literature data as determined from resistivity and magnetization \cite{RotterPhaseD}, specific heat \cite{Rotterspecificheat}, neutron scattering and magnetization measurements \cite{Avci} on polycrystalline materials. Thus determined $T_c (x)$ are in good agreement with each other and with measurements on single crystals \cite{Wencrystals,Liu}.
However, because the samples cut from various parts of the same crystal can have variation of $T_c$, we adopted the following approach in determining $x$ for each sample. We fitted the $T_c (x)$ determined by Avci et al., using two functions. In the underdoped regime we used a parabolic function, an approach frequently used in the cuprates \cite{Tallon}. For underdoped compositions this resulted in the formula $x= 0.40 -0.268*\sqrt{(1-T_c/39.5)}$, as shown by solid line in Fig.~\ref{f1_Tcx}. As can be seen from the comparison of the curve with actual data, the curve fits nicely the rapid rise of $T_c$ in the underdoped regime, the main interest in our study. Close to optimal doping the fit is not as good, the curve fails to reproduce the rather flat dependence in $x$ range from 0.3 to 0.5 \cite{Rotterspecificheat}.  For overdoped compositions the $T_c(x)$ is close to linear.

\begin{figure}[tbh]%
\centering
\includegraphics[width=8cm]{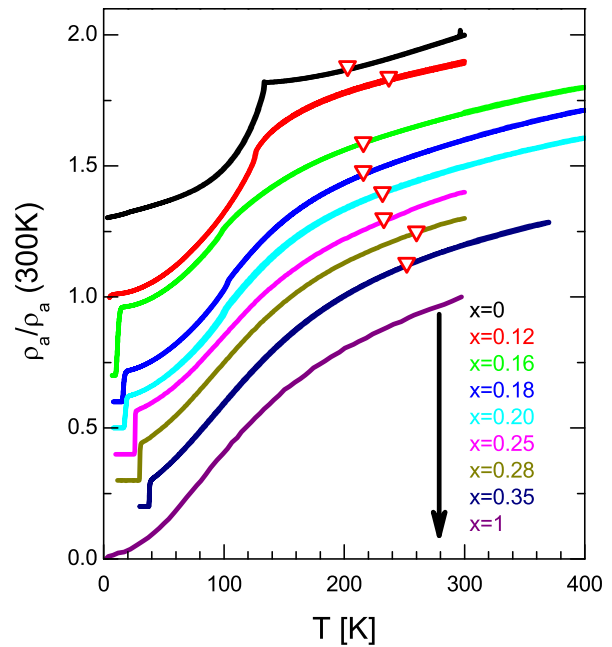}%
\caption{(Color online) Doping evolution of the temperature-dependent in-plane resistivity, $\rho_a(T)$, normalized to room temperature values, $\rho _a (300K)$. The curves are offset to avoid overlapping. Data for KFe$_2$As$_2$ are from Ref.~\onlinecite{Shiyan}. Down-triangles show the position of the maximum of the inter-plane resistivity.}
\label{f2_inplane}%
\end{figure}

Fig.~\ref{f2_inplane} shows temperature-dependent in-plane resistivity of BaK122 for several compositions from parent, $x$=0, to close to optimally doped, $x$=0.35. Several features should be noticed. First, the curves for samples with $x \leq 0.25$ ($T_c \leq $ 26K) show a small but clear anomaly on passing through the temperature of coupled structural/magnetic transition $T_{SM}$, with resistivity accelerating its decrease on cooling. The minimum distinct value of $T_{SM}$=70~K is observed in samples with $T_c$=26~K, and we could not resolve any feature in samples with $T_c \sim 30~K$. Both these observations are in good agreement with the position of the concentration boundary in neutron scattering data \cite{Avci}. In addition to a feature at $T_{SM}$, the in-plane resistivity shows a broad crossover with a slope change at a temperature of approximately 200~K, similar to previous measurements by Golubov et al. \cite{Zverev}. Similar crossover at about the same temperature is seen even in the terminal composition KFe$_2$As$_2$ \cite{Shiyan,Terashimaresistivity,HashimotoKpure} and in polycrystalline samples \cite{Rotter,Fukazawa}.

It is important to notice that the crossover feature is observed through all compositions from non-superconducting $x$=0.12 to heavily overdoped KFe$_2$As$_2$ ($x$=1), without any significant shift on a temperature scale. This insensitivity to doping is difficult to explain in any multi-band model, and by comparison with inter-plane resistivity, we suggest that this resistivity feature is a reflection of a pseudogap crossover at $T^*$ in the inter-plane transport, whose position is shown by arrows in Fig.~\ref{f2_inplane}. Another interesting observation is the very weak, if any, dependence of the value of room temperature resistivity on $x$. For all compositions the resistivity stayed close to 300 $\mu \Omega$ cm. This compares favorably to the values reported for KFe$_2$AS$_2$: 280 $\mu \Omega$ cm \cite{KAsfluxgrowth}, 360 $\mu \Omega$ cm \cite{Shiyan}, 480 $\mu \Omega$ cm \cite{Terashimaresistivity} and 600 $\mu \Omega$ cm \cite{HashimotoKpure}, with scatter presumably determined by cracks \cite{anisotropy} and inaccuracy of geometric factor determination.

\begin{figure}[tbh]%
\centering
\includegraphics[width=8cm]{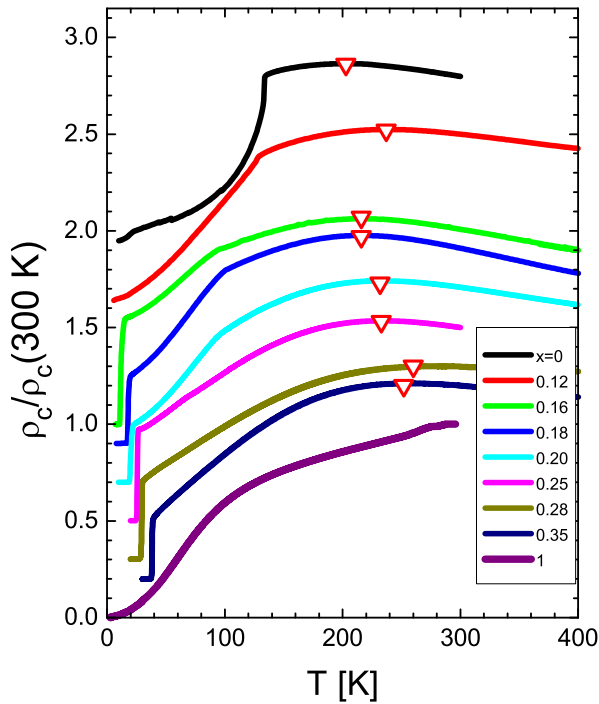}%
\caption{(Color online) Doping evolution of the temperature-dependent inter-plane resistivity, $\rho_c(T)$, normalized to room temperature values, $\rho _c (300K)$. The curves are offset to avoid overlapping. The data for KFe$_2$As$_2$ ($x$=1) are from Ref.~\onlinecite{TerashimacaxisPRL}. Down triangles show position of the $\rho _c (T)$ maximum, plotted in Fig.~\ref{f2_inplane} and in the phase diagram Fig.~\ref{f5_pseudogap}. }%
\label{f3_interplane}%
\end{figure}

\begin{figure}[tbh]%
\centering
\includegraphics[width=8cm]{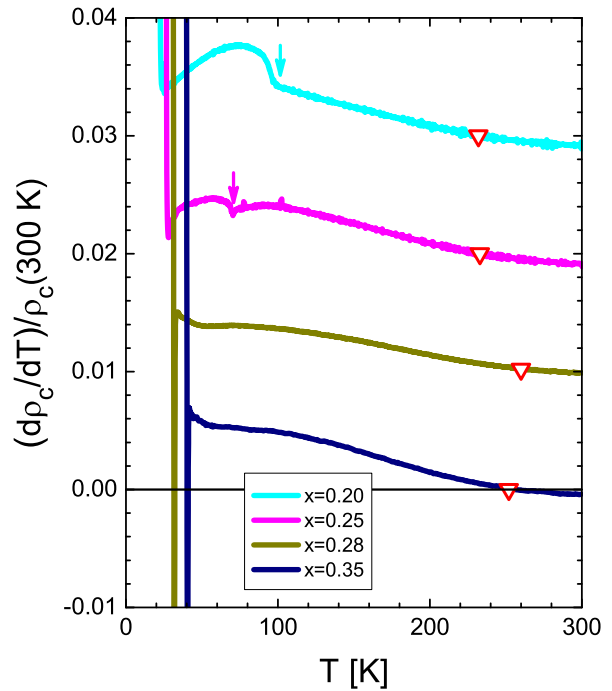}%
\caption{(Color online) Doping evolution of the temperature-dependent normalized resistance derivative, used for the determination of the structural/magnetic transition temperature $T_{SM}$ (arrows) in a doping range close to $x_{opt} \approx $0.4. The curves are offset to avoid overlapping. A small increase of the derivative preceding $T_c$ in samples $x$=0.28 and $x$=0.35 shifts up with the superconducting transition temperature and is most likely due to superconducting fluctuations. Down triangles show positions where the derivative is zero corresponding to the maximum of the inter-plane resistivity.
}%
\label{f4_interplanederivative}%
\end{figure}

In Fig.~\ref{f3_interplane} we plot the temperature-dependent inter-plane resistivity of BaK122. The feature due to the structural transition is much clearer in the $\rho _c(T)$, and we determine its position by plotting the derivative of the $\rho _c (T)$, as shown in Fig.~\ref{f4_interplanederivative}. In addition to the sharp feature in derivative in sample with $x$=0.25 at $T_{SM}$=70~K, a slight increase of the resistivity derivative is seen in samples with $x$=0.28 and $x$=0.35. This feature shifts to higher temperature with increasing superconducting $T_c$ and is most likely due to superconducting fluctuations. Note, that beyond this feature, samples with $x$=0.28 and $x$=0.35, close to optimal doping $x_{opt}$=0.4, show nearly linear temperature-dependent resistivity. This can be seen as a flattening of the resistivity derivative below $\sim$100~K above the superconducting transition.

\begin{figure}[tbh]%
\centering
\includegraphics[width=8cm]{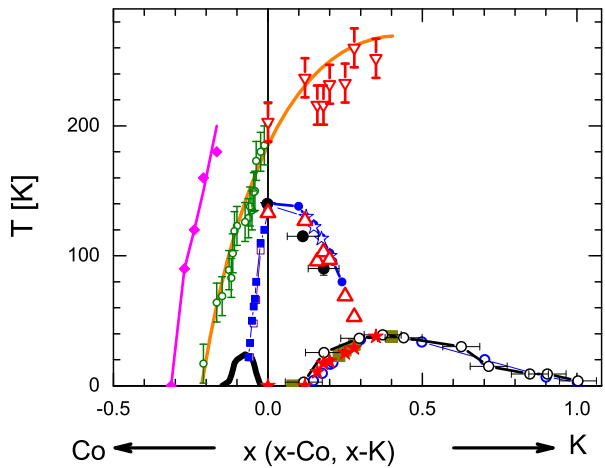}%
\caption{(Color online) Doping phase diagram of the hole-doped Ba$_{1-x}$K$_x$Fe$_2$As$_2$. The superconducting $T_c$ was determined from measurements on polycrystals (open circles),
Refs.~\onlinecite{RotterPhaseD,Rotterspecificheat,Avci}, see Fig.~\ref{f1_Tcx}. Dark-yellow solid squares show the data for single crystals, Ref.~\onlinecite{Wencrystals}, red stars show the $T_c(x)$ as determined in this study. The temperatures of the structural(solid blue dots) and magnetic (open blue stars) transitions are from neutron scattering, Ref.~\onlinecite{Avci} and x-ray (solid dots) study, Ref.~\onlinecite{Rotter}, red up-triangles show positions of the corresponding features in the inter-plane transport, Fig.~\ref{f3_interplane}. Red down-triangles show positions of maxima in the inter-plane resistivity at $T^*$.  For comparison we show the phase diagram for electron-doped Ba(Fe$_{1-x}$Co$_x$)$_2$As$_2$, Ref.~\onlinecite{pseudogap}.  Green and magenta lines represent positions of the maximum, $T^*$, and the minimum, $T_{CG}$, in temperature-dependent  inter-plane resistivity \cite{pseudogap}. The orange line is guide to the eyes. }%
\label{f5_pseudogap}%
\end{figure}

All these features in the temperature-dependent inter-plane resistivity are seen much below the temperature of the crossover point of resistivity from non-metallic behavior at high temperatures to metallic, with $\rho _c(T)$ decreasing on cooling, at low temperatures. The crossover temperature, $T^*$, shifts up very slightly with doping, from $\sim$200~K in parent compound to $\sim$250~K for BaK122 with $x$=0.35 (close to optimal doping). It is interesting that $T^*$ shows a strongly asymmetric doping dependence, with a rapid decrease in electron-doped BaCo122 and a slow rise in hole- doped BaK122, as summarized in the phase diagram plot of Fig.~\ref{f5_pseudogap}. Another distinct feature between the resistivity behavior of electron- and hole-doped materials is that the pseudogap affects both in-plane and inter-plane transport in hole-doped materials. The effect of $T^*$ on the in-plane transport is negligible in electron-doped materials.

\section{Discussion}

\subsection{Structural/magnetic ordering and inter-plane resistivity}

Contrary to BaCo122, magnetic ordering and the structural transition happen simultaneously in BaK122 at a temperature $T_s$=$T_N$ \cite{Avci}. Magnetic ordering should reconstruct the Fermi surface, opening a nesting or superzone gaps on electron and hole pockets. In hole-doped materials this gap opening, instead of leading to a resistivity increase, leads to an accelerated resistivity decrease (increase of resistivity derivative), suggesting that the main effect comes from a change in the inelastic scattering due to taming down the contribution of pre-transition  fluctuations of the order parameter. The parts of the Fermi surface which are not affected by the SDW gap \cite{MazinSDW}, enjoy a notably reduced inelastic scattering in the magnetically ordered phase \cite{HallWen,AlloulPRL,detwinning}. In the parent compound, in which disorder, introduced by dopants, is absent and thus elastic scattering is small, this decrease of scattering overcomes the loss of the carrier density so that the total conductivity increases below $T_{SM}$. Since the inter-plane transport is dominated by the most warped parts of the Fermi surface \cite{anisotropy}, least affected by the SDW super-zone gap, the inter-plane resistivity should be affected much less by the SDW gap opening than $\rho _a$. This is indeed seen, in BaK122, very similar to BaCo122.

\subsection{Maximum in the temperature-dependent inter-plane resistivity at $T^*$}

In BaCo122 the decrease of the inter-plane resistivity below $T^*$ shows a clear correlation with the NMR Knight shift, which was the reason for our suggestion of its relation to a pseudogap. At temperatures below $T^*$  both the Knight shift and the inter-plane resistivity in BaCo122 follow the expectations of a metal with temperature-independent density of states, but become temperature-dependent at $T>T^*$. Very recently similar measurements have been undertaken in optimally doped BaK122 \cite{NMRoptimal}, and in stoichiometric KFe$_2$As$_2$ \cite{NMRK122}. At optimal doping the NMR Knight shift is increasing with temperature while the spin-relaxation rate, 1/$T_1T$, decreases and becomes constant above $\sim$200~K. In overdoped KFe$_2$As$_2$ the Knight shift was found to be nearly temperature independent, however, the temperature dependence of the Korringa ratio indicated strong electron correlations.
We need to notice though, that the decrease of the inter-plane resistivity, despite being very small, would be very difficult to explain by only a change of the scattering mechanism. It would require activation of carriers by excitations over the partial gap on the Fermi surface (pseudogap).

\section{Conclusions}

Measurements of the inter-plane resistivity in BaK122 show that the magnetic/structural transition does not lead to resistivity increase, i.e. the loss of the carrier density at the transition is insignificant. On transition suppression with doping, temperature-dependent inter-plane resistivity becomes nearly linear above the superconducting transition, suggesting the existence of a quantum critical point. In addition to a feature in the temperature-dependent resistivity at $T_{SM}$, a broad maximum is found at a temperature $T^* \sim$200~K, which by comparison with the electron-doped BaCo122 \cite{pseudogap}, we assign to the formation of pseudogap.

Despite the effect of doping in multi-band metallic system may be quite complicated, comparison of the hole-doped BaK122 with the electron-doped BaCo122 shows significant difference. A pseudogap resistive crossover at $T^*$ in the inter-plane resistivity vanishes with doping in BaCo122 but remains intact in BaK122. The crossover affects temperature-dependent in-plane resistivity in BaK122, however, it does not in BaCo122.

On Co-doping the in-plane resistivity at room temperature decreases by approximately a factor of 3 from $x$=0 to $x_{Co}$=0.31, while it does not show much variation on hole doping. This result may suggest that inelastic scattering is strongly suppressed in the case of Co-doping, while it remains strong for hole-doped compositions. Since a similar trend is seen for magnetic fluctuations, it is natural to expect that magnetic scattering contributes significantly to normal state inelastic scattering. Supporting this conclusion, BaCo122 doped beyond the superconducting dome shows simple metallic behavior in resistivity (both in- and inter-plane) and Pauli susceptibility
 \cite{pseudogap}, and temperature-independent Hall constant  \cite{Halljapanese,HallWen}, with all pseudogap features suppressed by $x_{Co}$=0.31. On the other hand, anomalous temperature-dependent resistivity with the resistivity crossover, suggesting the existence of the pseudogap, is preserved in KFe$_2$As$_2$.

Finally we would like to point to a certain similarity in the critical behavior of the inter-plane resistivity in BaK122 and in CeCoIn$_5$. In CeCoIn$_5$, a true critical behavior at a field-tuned QCP \cite{Paglione,Bianchi} with $T$-linear resistivity and violation of the Wiedemann-Franz law is observed for transport along the tetragonal $c$-axis \cite{Science}, while transport along the plane obeys the Wiedemann-Franz law \cite{nonvanishing}. This is similar to the difference in the temperature dependence of $\rho_a(T)$ and $\rho_c(T)$ in BaK122, with the latter being more linear at optimal doping.

\section{Acknowledgements}

We thank T. Terashima and M. Kimata for sharing resistivity data files, and Y. Liu and C. T. Lin for providing unpublished doping-dependent $T_c$ data. Work at the Ames Laboratory was supported by the Department of Energy-Basic Energy Sciences under Contract No. DE-AC02-07CH11358. Work in China was supported by the Ministry of Science and Technology of China, project 2011CBA00102.


\end{document}